# Prediction of the Al-rich part of the Al-Cu phase diagram using cluster expansion and statistical mechanics

S. Liu[a], E. Martínez[b], J. LLorca[a,c,1]

[a] IMDEA Materials Institute, C/Eric Kandel 2, Getafe 28906 – Madrid, Spain
[b] Theoretical division, T-1, Los Alamos National Laboratory, Los Alamos 87545 NM, USA.
[c] Department of Materials Science. Polytechnic University of Madrid. E. T. S. de Ingenieros de Caminos. 28040 – Madrid, Spain

**Abstract**

The thermodynamic properties of α-Al and other phases (GP zones, θ", θ' and θ) in the Al-rich part of the Al-Cu system have been obtained by means of the cluster expansion formalism in combination with statistical mechanics. This information was used to build the Al-rich part of the Al-Cu phase-diagram taking into account vibrational entropic contributions for θ', as those of the other phases were negligible. The simulation predictions of the phase boundaries between α-Al and either θ", θ' or θ phases as a function of temperature are in good agreement with experimental data and extend the phase boundaries to a wider temperature range. The DFT calculations reveal the presence of a number of metastable Guinier-Preston-zone type configurations that may coexist with α-Al and θ" at low temperatures. They also demonstrate that θ' is the stable phase below 550K but it is replaced by θ above this temperature due to the vibrational entropic contribution to the Gibbs energy of θ'. This work shows how the combination of cluster expansion and statistical mechanics can be used to expand our knowledge of the phase diagram of metallic alloys and to provide Gibbs free energies of different phases that can be used as input in mesoscale simulations of precipitation.

.

Keywords: Cluster expansion, Al-Cu, statistical mechanics, Phase diagram, precipitation.

---

[1] Corresponding Author.
Email address: javier.llorca@imdea.org (J. LLorca)

# 1. Introduction

The application of advanced simulations to design novel structural alloys has progressed rapidly in recent years owing to several factors which include the ever increasing computational power, the maturity of the simulation tools at different length scales and the emerging of novel multiscale modelling strategies to bridge length and time scales [1-2]. Nevertheless, the first building blocks of these multiscale simulations for alloy design are the phase diagrams that provide information about the different stable thermodynamic phases as a function of the alloy composition and temperature. The standard strategy to predict the phase diagram of an alloy system is based in the CalPhad (computation of phase diagrams) methodology [3] which uses approximate analytical expressions of the Gibbs free energy of the different phases with adjustable parameters that are optimized from the available experimental and theoretical results. This information is used to predict the stable phases and their thermodynamic properties in regions without experimental information.

Although this methodology is widely used in industry and academia, it is well understood that the accuracy of the predictions is hindered by the limitations of the theoretical approach as well as by the lack of reliable experimental data. For instance, there are still uncertainties and inconsistencies in the current accepted version of the Al-Cu phase diagram [4-5], although this phase diagram has been studied for one hundred years due to its huge technological interest. These limitations are due to the presence of complicated order-disorder transitions as well as to the rich variety of metastable and stable phases, whose solubility limits are very difficult to determine experimentally [6]. Cu can be found as the one major alloying elements in most of wrought Al alloys that are widely used in transportation (aerospace, automotive, railways, marine, etc.) because of the low density, limited cost, ease of fabrication and excellent combination of mechanical properties (strength, ductility and toughness) of these alloys [7]. Besides solid solution strengthening, the presence of Cu increases the strength of Al by the formation of a number of metastable and stable phases that precipitate from the supersaturated solid solution during heat treatments at different temperatures, namely Guinier-Preston (GP) zones, $\theta''$ ($Al_3Cu$), $\theta'$ ($Al_2Cu$) and $\theta$ ($Al_2Cu$) [8]. Nevertheless, the stability curves of the different metastable phases as a function of temperature are difficult to determine using the CalPhad methodology and the available experimental data are very limited. For instance, the only data of the solvus temperature of $\theta''$ and $\theta'$ come from a few experiments carried out many years ago in which the presence of the phases was ascertained by indirect methods (reversion hardness, electrical resistivity) but direct evidence of the precipitates was not ascertained by means of electron microscopy [4-5] and the situation is similar in the case of the phase boundary between α-Al and θ precipitates when the Cu content is close to 0.33. In fact, the only phase boundary well-established is the solvus line of θ precipitates [4].

A different approach to determine the phase diagram of alloys is based on the cluster expansion (CE) formalism [9-11], which is able to represent the thermodynamic properties of a multicomponent system as a linear series of cluster basis functions multiplied by constant expansion coefficients that depend on the chemistry and crystal structure of the system. The coefficients of the CE can be determined from first-principles calculations of different configurations and this energies provided by the CE formalism are used in combination with statistical mechanics simulations to access the thermodynamic quantities, enabling parameter-free predictions of the phase diagrams of metallic systems, such as Ni-Rh [12], Fe-Ni [13], Cu-Pd [14], Al-Sc [15] and Mg-Nd [16] including metastable phases [17].

It is obvious that the combination of traditional methods to determine phase diagrams with CE approaches can be extremely useful to eliminate the uncertainties induced by experimental errors and also to provide quantities difficult to obtain experimentally, such as the Gibbs free energies of metastable phases which are -however- very important to determine the nucleation and growth of precipitates during thermomechanical treatments [18-19]. Within this framework, the thermodynamic properties of the most relevant phases in the Al-rich part of the Al-Cu phase diagram (α-Al, θ'', θ' and θ) are determined as a function of temperature and composition using the CE formalism. The contribution of the vibrational entropy is also included in the analysis in addition to the configurational entropy provided by the CE formalism because the stability of the θ' phase is known to be affected by this quantity [19-21]. This information is used to build the Al-rich region of the Al-Cu phase diagram, which is compared with the limited experimental data available in the literature for the metastable phases. In addition, the thermodynamics of the Guinier-Preston (GP) zones at the early stage of phase separation is also analyzed. Our results show how the combination of the CE methodology and statistical mechanics can be used to expand our knowledge of the phase diagram of metallic alloys and to provide Gibbs free energies of different phases that can be used as input in mesoscale simulations of precipitation.

## 2. Theoretical background

The theoretical background to obtain the thermodynamic properties of the binary alloy is briefly recalled here for the sake of completion. The formation energy per atom of an $Al_{1-x}Cu_x$ binary crystal with a certain crystalline structure in the athermal limit can be expressed as

$$E^f(Al_{1-x}Cu_x) = E(Al_{1-x}Cu_x) - (1-x)E(Al) - xE(Cu) \qquad (1)$$

where $E(Al)$ and $E(Cu)$ stand for the relaxed energies per atom of pure Al and pure Cu with the same crystalline structure and $E(Al_{1-x}Cu_x)$ is the relaxed energy per atom of the $Al_{1-x}Cu_x$ crystal. The relaxed energies -and, thus, the formation energy- can be determined using first principles calculations based on the Density Functional Theory (DFT) in which both atomic coordinates and lattice vectors are allowed to relax at pressure P=0. For each composition, the formation energy will depend on the actual configuration, i.e. the position of the Al and Cu atoms in the lattice, leading to a formation energy distribution. The configurations in the convex hull have the minimum formation energy per atom for a given composition, as compared with other configurations, and stand for the stable equilibrium phases at 0K. The main parameter governing the instability of a configuration $s$ is the distance to the convex hull $d_s(\vec{\sigma})$, where $\vec{\sigma} = (\sigma_1, \sigma_2, ... \sigma_N)$ is the discrete configurational variable that represents the configuration of the crystal with $N$ sites. In the case of binary alloy, $\sigma_i = +1$ if the site $i$ is occupied by one of the chemical species (Al) and $\sigma_i = -1$ otherwise (Cu).

The CE formalism shows that the formation energy per atom can be expressed as a function of the configuration according to [9-11]

$$E^f(\vec{\sigma}) = \sum_f V_f \prod_{i \in f} \sigma_i \qquad (2)$$

where $V_f$ stand for the effective cluster interaction (ECI) coefficients and $\prod_{i \in f} \sigma_i$ are a specific set of crystal basis functions that describe the different types of interactions (pairs, triplets, quadruplets, etc.) in the system. For a binary alloy crystal with N sites in the lattice, there are $2^N$ different crystal basis functions since each lattice site can be occupied by either chemical species but the CE model usually converges rapidly and only a relatively few number of interactions are needed to accurately calculate the formation energy of a given configuration. The ECI coefficients can be determined from the formation energies obtained by DFT for a number of configurations in the system using different approaches.

The equilibrium thermodynamic properties of an ergodic system can be obtained from the partition function Z. The semi-grand-canonical ensemble is normally used in alloys to compute phase boundaries. It is characterized by a fixed number of sites N in given crystal lattice, constant temperature T and chemical potential μ according to [22]

$$Z = \sum_s e^{-\beta(E_s^f - \Delta \mu x)N} \qquad (3)$$

where $\beta = 1/k_b T$, $k_b$ the Boltzmann constant, $\Delta \mu = \mu_{Cu} - \mu_{Al}$ is the difference in chemical potential between the two species and $x$ is the composition (expressed in this case as the fraction of Cu in the system). The sum in $s$ extends for all possible states of the system. The CE formalism provides an efficient tool to determine the formation energy $E_s^f$ of crystals with different configurations while the partition function can be evaluated using the metropolis Monte Carlo (MC) method [23].

The connection between the thermodynamic grand potential Φ and the partition function is given by [22]

$$\beta \Phi = -\ln Z \quad (4)$$

where

$$\Phi = U - TS - \Delta\mu x \quad (5)$$

where $U$ and $S$ stand for the internal energy and entropy of the system. In the case of solid state transformations, the difference between the Gibbs and Helmholtz free energies can be neglected at atmospheric pressure because the product of pressure $P$ with changes in specific volume $(\Delta V)$ are small, and the Gibbs free energy for each ground state phase can be obtained from the grand potential according to

$$G = \Phi + \Delta\mu x \quad (6)$$

The determination of the phase transition boundaries between two phases depends on their lattice structure. If they have the same lattice structure, i.e. their thermodynamics properties were obtained from the same CE, the phase transition boundaries are determined by $\Phi$. The stable phase for any value of μ and T is the one with the lowest $\Phi$ and the phase transition boundaries are directly determined by the intersection of $\Phi$ for both phases [24]. However, the grand potentials should be used with care to compare phases with different lattice structures since the reference chemical potential must be equal, which makes the thermodynamic integration cumbersome. Hence, the phase transition boundaries are determined from the common tangent between the Gibbs free energies of both phases. We should note here that the reference values of $G$ are different for different structures and they should be transferred to the same reference before comparison, as it will be explained in Section 4.4.

## 3. Simulation details

### 3.1 DFT calculations

The CASM (Clusters Approach to Statistical Mechanics) code was used to generate symmetrically distinct crystal configurations [25]. The relaxed energies of the configurations were obtained from DFT calculations using Quantum Espresso [26] in the ultra-soft pseudopotential mode. The exchange-correlation energy was evaluated using the Perdew-Burke-Erzenhof approach with 53 Ry as the energy cut-off [27]. The Brillouin zone was sampled using a 23×23×23 Monkhorst-Pack grid in a fcc Al primitive cell of one atom, a 14×14×14 grid in a bct (θ') primitive cell of three atoms and a 12×11×11 grid in a bct (θ) primitive cell of six atoms. K-point grids differ depending on the unit cell of each configuration and CASM keeps the mesh density constant for all configurations. The atomic positions, lattice parameters and angles were allowed to relaxed at P=0 for each structure.

In the case of intermetallic compounds, it is important to take into account the vibrational entropic contribution to the free energy. The phonon contribution to the free energy can be accounted for by means of the quasi-harmonic approximation [28], where the thermal properties of solid materials are traced back to those of a system of non-interacting phonons whose frequencies are, however, allowed to depend on volume or on other thermodynamic constraints. Within this framework, the vibrational entropic contributions for each phase (fcc Al and Cu as well as θ", θ', and θ) were calculated from the phonon density of states $g(\omega)$, were $\omega$ denotes the volume dependent phonon frequencies, according to [29]:

$$S_v(T) = k_b \int_0^\infty \frac{\frac{\hbar\omega}{k_bT}}{\exp\left(\frac{\hbar\omega}{k_bT}\right)-1} g(\omega)d\omega - k_b \int_0^\infty g(\omega) \ln[1 - \exp\left(\frac{\hbar\omega}{k_bT}\right)]d\omega \quad (7)$$

where ℏ is the reduced Planck's constant. The phonon density of states was determined by finite displacement method [29], which is also known as supercell method. 3×3×3 supercells for Al and Cu, 3×3×2 supercells for θ" and θ', and 2×2×3 supercells for θ were used.

### 3.2 Cluster expansion formalization

In order to determine the ECIs for each lattice structure, an initial atomic spacing was set for each type of cluster interaction (pair, triplets, etc.) and the corresponding clusters within this atomic spacing were used to build the CE. In order to reduce the computational cost during the MC simulation, it is important to limit the sets of clusters included in the CE while enough accuracy is retained. The optimum set of clusters and the corresponding ECIs for each lattice structure were determined following the strategy detailed in [24] and implemented in CASM [25]. To this end, the data of the formation energies for each lattice was divided in 10 groups. Each group was for testing the accuracy of CE in turn, while remaining 9 groups were used for training. The optimum ECIs in each iteration were determined using a genetic algorithm [25] and the estimator of the accuracy in all cases was a cross-validation score based on a weighted least-squares fit to obtain better predictions for the structures with low energies. Thus, the distance to the convex hull of each structure, $d(\vec{\sigma})$, was modified according to $w(\vec{\sigma}) = A \exp(d(\vec{\sigma})/k_bT) + B$, where A+B is the maximum weight given to the ground states on the convex hull and B is the minimum weight given to the microstates far from the hull. We used the values A+B=20, B=1, $k_bT = 0.01$ eV/atom.

### 3.3 Calculation of the grand potentials and Gibbs free energies

A combination of low temperature expansion [30] of the free energy with metropolis MC simulations was used to determine the grand potentials and the Gibbs free energies of the different structures. In the two-phase region, where the $\Phi$ of two phases are equal, the difference in the Gibbs free energy between both phases is given, according to eq. (6), by $\Delta G = \Delta U - T\Delta S = \Delta\mu x$. At low temperature, where the adjacent stable phases are stoichiometric at their ground state, the slope of the line segment connecting them in the convex hull plot plays the role of the difference in chemical potential $\Delta\mu$ because the internal energy is directly given by the formation energies on the convex hull, and the entropy term $T\Delta S$ can be neglected. Which two phases coexist at low temperature can be selected from the difference in chemical potential $\Delta\mu$. Therefore, a low temperature expansion was first run at 100K over a large range of $\Delta\mu$, to obtain the stability range of the difference in chemical potential of each pair of phases. Moreover, the grand potential reference for each phase was also obtained from CASM using the low temperature expansion [22, 30].

At high temperature, the grand potential of each phase was calculated using the metropolis MC algorithm in CASM. From eqs. (3) and (4), the grand potential $\Phi$ for each phase can be defined by the following total differential

$$d(\beta\Phi) = N(E_s^f - \Delta\mu x)d\beta - N\beta x\, d\Delta\mu \tag{8}$$

and $\Phi$ can be obtained for a given $\Delta\mu$ as

$$\beta^{end}\Phi(\beta^{end}, \Delta\mu) = \beta^{begin}\Phi(\beta^{begin}, \Delta\mu) + N \int_{\beta^{begin}}^{\beta^{end}} \langle E_s^f - \Delta\mu x \rangle d\beta \tag{9}$$

and for a given $\beta$ as

$$\Phi(\beta, \Delta\mu^{end}) = \Phi(\beta, \Delta\mu^{begin}) - N \int_{\Delta\mu^{begin}}^{\Delta\mu^{end}} \langle x \rangle d\Delta\mu \tag{10}$$

where $\langle E_s^f - \Delta\mu x \rangle$ and $\langle x \rangle$ stand for the ensemble averages and $\beta^{begin}$ and $\beta^{end}$ and $\Delta\mu^{begin}$ and $\Delta\mu^{end}$ stand for the range of temperatures and chemical potentials explored in the MC simulation [22]. The grand potentials for each phase obtained from low temperature expansion at 100K were used as starting points for the integration. The fine-grid metropolis MC calculations were first run by increasing and decreasing temperature at increments of 10K over the range of 100K ≤ T ≤ 900K at each chemical potential. The range of chemical potentials explored for each pair of phases was larger than the stability range of each pair of phases to include the phase boundaries. Afterwards, the metropolis MC simulations were carried out by increasing and decreasing the chemical potential at increments 0.05 eV at each temperature to approach the two-phase region[1]. Once the grand potential for each phase was determined, the Gibbs free energy can be obtained from eq. (6). The MC calculations were performed in periodic supercells of dimensions 10×10×10 primitive unit cells. For each value of T and Δμ, a MC calculation was performed consisting of a number of equilibrating passes till the precision of the sampling properties reached 95%, followed by 1000 passes for calculating the thermodynamic averages. A pass is defined as $N_{sites}$ attempted flips, $N_{sites}$ being the number of sites in the Monte Carlo cell with variable occupations.

---

[1] It should be noted that the fcc system contains one atom per unit cell, the bct (θ') system 3 atoms per unit cell and the bct (θ') 6 atoms per unit cell.

## 4. Results

### 4.1 Structures of interest

The available experimental information indicates the different phases that can be found in the Al-rich part of the Al-Cu system after precipitation from the supersaturated solid solution, namely GP zones, θ" ($Al_3Cu$), θ' ($Al_2Cu$) and θ ($Al_2Cu$) [5-7, 31-32]. The GP zones are layers of Cu atoms parallel to the {001} planes of fcc α-Al lattice. The structures of α-Al, θ", θ' and θ are depicted in Fig. 1. α-Al has a fcc structure while θ" shows a face-centered tetragonal (fct) structure, and the height of the unit cell is nearly two times that of α-Al. θ' shows a body-centered tetragonal (bct) structure and $a_{θ'}$ is the same as $a_{α-Al}$, while $c_{θ'}$ is much lower than $2a_α$. Finally, θ shows another bct structure with lower symmetry.

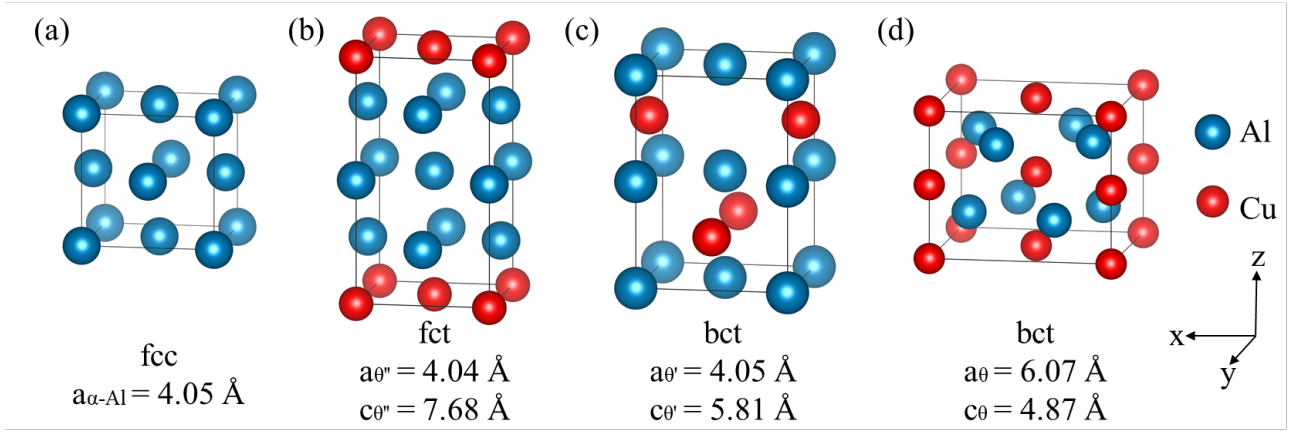

**Fig. 1** Crystal structures of different phases in the Al-rich part of the Al-Cu system. (a) α-Al. (b) θ" ($Al_3Cu$). (c) θ' ($Al_2Cu$) and (d) θ ($Al_2Cu$). Cu atoms are red and Al atoms blue.

### 4.2 Vibrational entropic contribution

The vibrational entropic contribution to the free energy is given by [19]:

$$E_v = -TS_v(T) \qquad (11)$$

where $S_v(T)$ is the vibrational entropy given by eq. (7). Thus, the contribution of the vibrational entropy to the formation energy, $E_v^f$, of θ", θ' and θ compounds with stoichiometry $Al_{1-x}Cu_x$ can be expressed as:

$$E_v^f(Al_{1-x}Cu_x) = E_v(Al_{1-x}Cu_x) - (1-x)E_v(Al) - xE_v(Cu) \tag{12}$$

According to the calculated phonon density of states, the formation energies of each phase including the vibrational entropic contribution are compared in Fig. 2. It shows that the vibrational entropy of θ' is important while those of θ" and θ can be neglected, which is consistent with the results of Liu *et al.* [19]. Thus, only the vibrational entropy of θ' was considered in this paper.

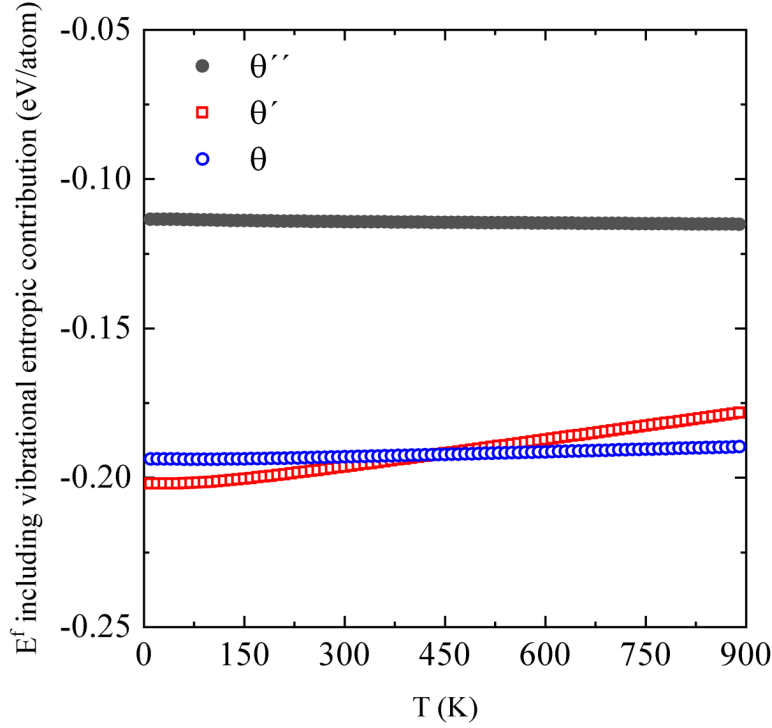

**Fig. 2** Formation energies of θ", θ' and θ as a function of temperature including the vibrational entropic contribution.

### 4.3 Phase boundary between α-Al and θ"

θ" is an ordered phase with crystalline structure similar to α-Al in which some Al sites have been replaced by Cu atoms, although the atomic positions in θ" differ somewhat from those of α-Al because of the presence of the Cu atoms. Therefore, the fcc Al primitive cell was utilized as the motif structure, and 343 symmetrically distinct configurations were generated with randomly arranged Al and Cu atoms on the lattice sites up to 12 atoms per unit cell. The formation energies of the configurations obtained by DFT are shown in Fig. 3. They indicate that θ" is the only ground-state phase in the Al-rich side of the fcc Al-Cu system. Several additional ground-state structures can be identified in the Cu-rich side besides the known ground states such as $\eta_2$ (AlCu) [33] and δ ($Al_2Cu_3$) [34] with fcc structure. The structures of the ground state phases are shown in Fig. 4. All phases have (001) planes ordering up to *x*=0.66, so that the Cu(001) and Al(001) layers are staggered along the [001] direction. The sequence is Al-Al-Al-Cu in θ" ($Al_3Cu$), Al-Cu-Al-Cu-Cu in $Al_2Cu_3$ and Al-Cu-Cu in $AlCu_2$. The structure of AlCu3 consists of alternating Cu(001) and Al & Cu(001) layers. The AlCu phase with alternative Cu(001) and Al(001) layers lies slightly above the convex hull and it is not stable at 0K although it is reported to be a high-temperature phase in the Al-Cu system [33]. We should note that there are many other configurations close to the segment connecting α-Al and θ" in Fig. 3. These structures are shown in Fig. 5: they all have (001) planes ordering and Cu(001) are intercalated between Al(001) layers. Nevertheless, structures with consecutive Cu(001) were never found near the convex hull.

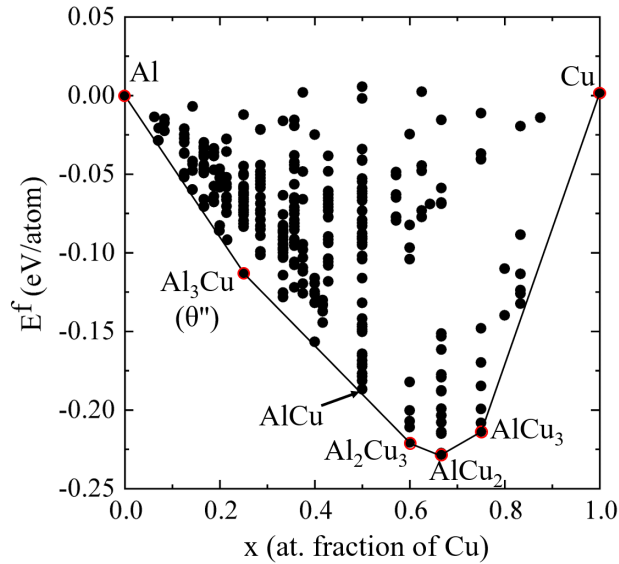

**Fig. 3** Formation energies of different configurations in the fcc Al-Cu system calculated by DFT. The ground state phases in the convex hull are marked with a red circle.

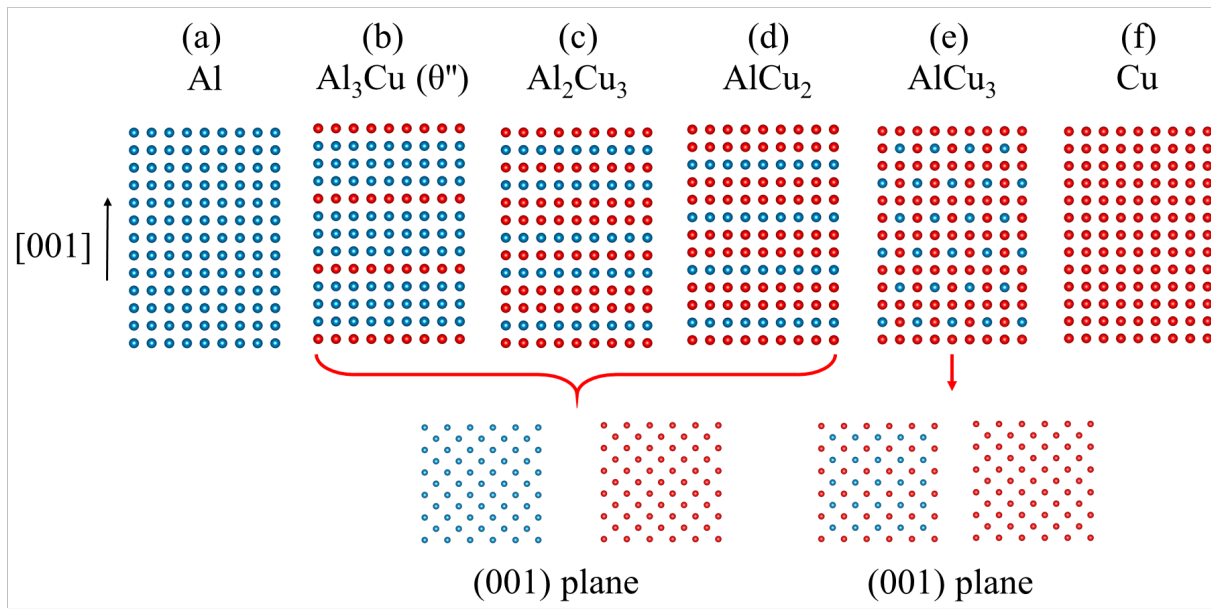

**Fig. 4** Ground-state ordered phases in the fcc Al-Cu system. Cu atoms are red and Al blue.

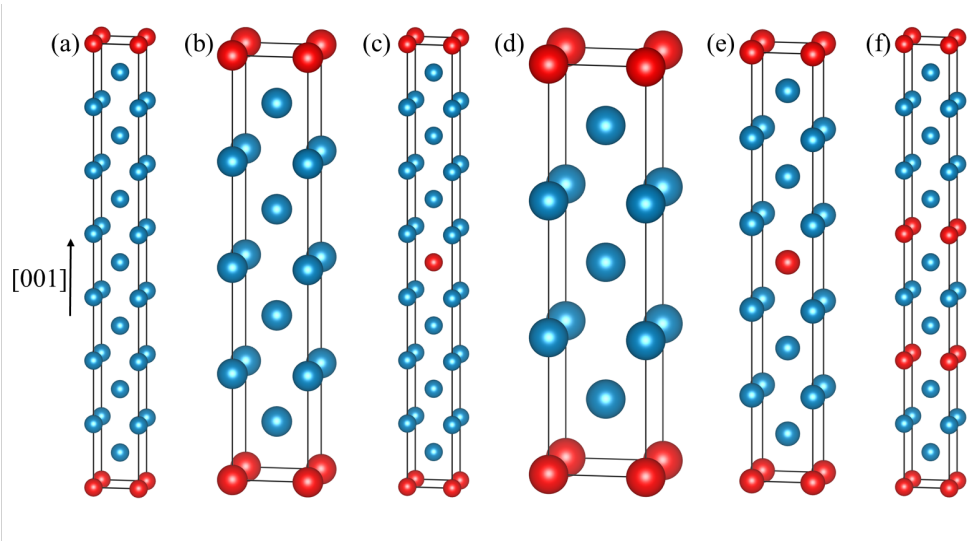

**Fig. 5** Configurations close to the line segment connecting α-Al and θ″ in Fig. 3 with different Cu content. (a) *x*=0.071. (b) *x*=0.125. (c) *x*=0.143. (d) *x*=0.167. (e) *x*=0.2. (f) *x*=0.214. Cu atoms are red and Al blue.

The clusters of atoms with a maximum atomic spacing of 12 Å for pairs, 7 Å for triplets, and 5 Å for quadruplets were considered to fit the ECI coefficients for the fcc Al-Cu system. According to the crystallography of the fcc structure and the atomic spacing of Al primitive cell, 60 clusters were initially included. The final optimized ECIs set for the fcc Al-Cu system includes an empty cluster interaction, a point cluster interaction, 5 pair interactions, 5 triplet interactions and a quadruplet interaction, and their values are presented in Table S1 in the Supplementary Material. The corresponding cross-validation score of the least-squares fitting of the CE was only 0.01 eV/atom.

The phase boundaries between α-Al and θ″ were determined as follows. Low thermal expansion and MC simulations were performed to obtain $\Phi(\Delta\mu, T)$ of α-Al and θ″ as a function of temperature and difference in chemical potential as explained in Section 3.3 using CASM. At a given temperature, the stable phase for each chemical potential can be obtained by comparing $\Phi^{\alpha-Al}$ and $\Phi^{\theta''}$, as shown in Fig. 6a for T = 700K. The black curve in Fig. 6a is calculated from the ground state α-Al by increasing $\Delta\mu$ while the red curve is calculated from θ″ by decreasing $\Delta\mu$. The intersection $\Phi^{\alpha-Al} = \Phi^{\theta''}$ corresponds to $\Delta\mu$ where α-Al and θ″ coexist. Because the composition is conjugated to the difference in chemical potential, the intersection point can be mapped into the relationship between them, which is plotted in Fig. 6b. The region in which $\Delta\mu$ is constant while the composition varies sharply from x= 0.045% to 0.26% in Fig. 6b corresponds to the two-phase region at 700K. Only one phase is stable below and above this value of $\Delta\mu$ at this temperature. It should be noted that the composition-chemical potential curves calculated from α-Al by increasing $\Delta\mu$ and from θ″ by decreasing $\Delta\mu$ in Fig. 6b are not coincident due to the hysteresis phenomenon during phase transition [35] and they cannot be used to determine $\Delta\mu$ in the two-phase region.

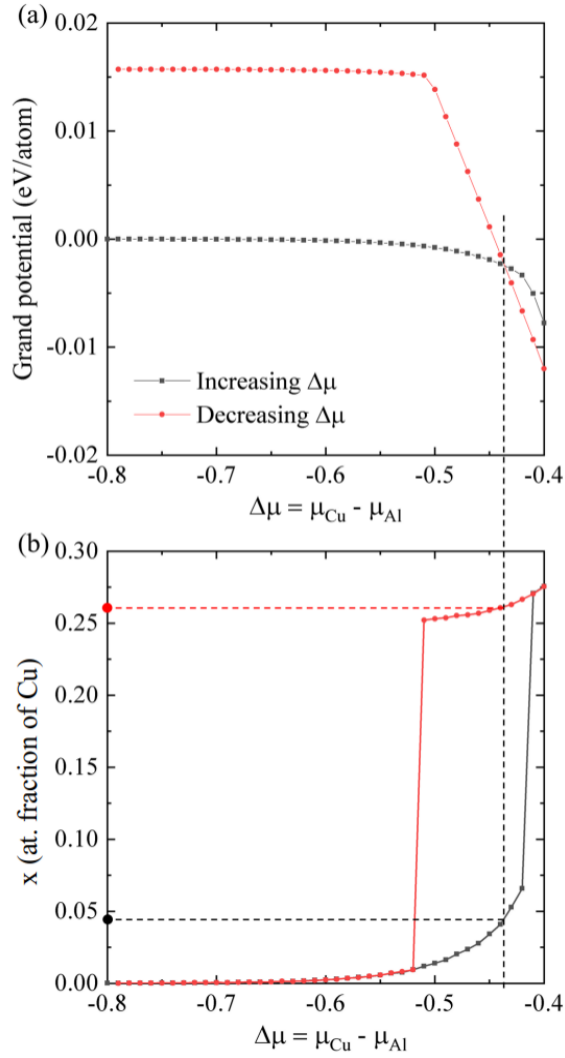

**Fig. 6** (a) Calculated grand potential $\Phi$ as a function of the difference in chemical potential $\Delta\mu$ in the α-Al and θ" region at 700K. (b) Calculated composition $x$ as a function of the difference in chemical potential $\Delta\mu$ in the α-Al and θ" region at 700K.

This procedure can be used at different temperatures to determine the phase transition boundary between α-Al and θ", which is shown in Fig. 7, and compared with the limited experimental data available in the literature from hardness reversion experiments [36-37] as well as resistivity measurements [37]. They are limited to a small range of Cu content ($x < 0.02$) and, in addition, it should be noted that these data are approximate because the reversing on temperature (used to determine the solvus line in Fig. 7) depends on the size of the precipitates and, thus, on the prior aging conditions. On the contrary, the predictions obtained from the cluster expansion provide the boundaries of the two-phase region at any temperature and they can also be used to predict the Gibbs free energy of the θ" precipitates as a function of the Cu content. This information is necessary to determine the dynamics of precipitate growth using phase field models [19].

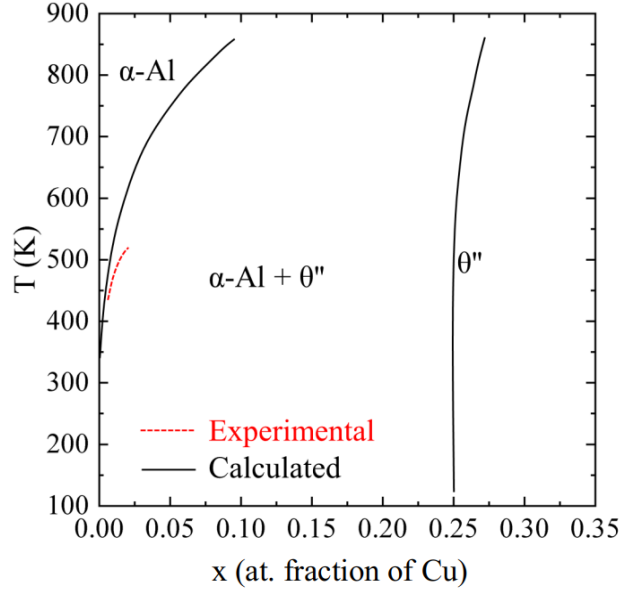

**Fig. 7** Calculated and experimental [37-38] phase boundaries between α-Al and θ".

**4.4 Phase boundary between α-Al and θ'**

The primitive cell of bct θ' was used as the motif structure to get the thermodynamic properties of θ' ($Al_2Cu$). 100 symmetrically distinct configurations were generated with randomly arranged Al and Cu atoms on the lattice sites with up to 9 atoms per unit cell, but 9 of them over-relaxed to very different types of lattice. If these over-relaxed configurations to construct the CE, the convergence and accuracy of the predictions decreased and even the ground states on the convex hull could not be predicted accurately, in agreement with previous results [39-40]. Therefore, these 9 configurations that over-relaxed to different lattice structures were not used to build the CE. The formation energies of the 91 configurations calculated by DFT that were used to build the CE are shown in Fig. 8a. Clusters of atoms with maximum atomic spacing of 10 Å for pairs, 5 Å for triplets, and 5 Å for quadruplets were considered to determine the ECI coefficients of the CE for the bct (θ') Al-Cu system with a total of 101 clusters. The optimized ECI coefficients set includes 2 point cluster interactions, 4 pair interactions, 2 triplet interactions and 6 quadruplet interaction, which are detailed in Table S2 in the Supplementary Material). The corresponding cross-validation score of all configurations was 0.06 eV/atom, while that of the configurations with composition x⩽0.33 was 0.03 eV/atom.

MC simulations were performed to get the Φ of θ' as a function of temperature and the difference of chemical potential and, therefore, $G^{θ'}$ was calculated as a function of composition at each temperature according to eq. (6). However, the reference values of $G^{θ'}$ were calculated with respect to energies of pure Al and pure Cu with bct (θ') structure, which are different from those of the stable phases fcc Al and fcc Cu. As the formation energy of each configuration only depends on the composition according to eq. (1), the formation energy $E^f$ can be changed from the reference to the bct structure in Fig. 8 to the reference in the fcc structure by adding an energy difference $\Delta E^f$ that only depends on composition. The same energy difference $\Delta E^f$ should be added to $G^{\theta'}$ according to eq. (6).

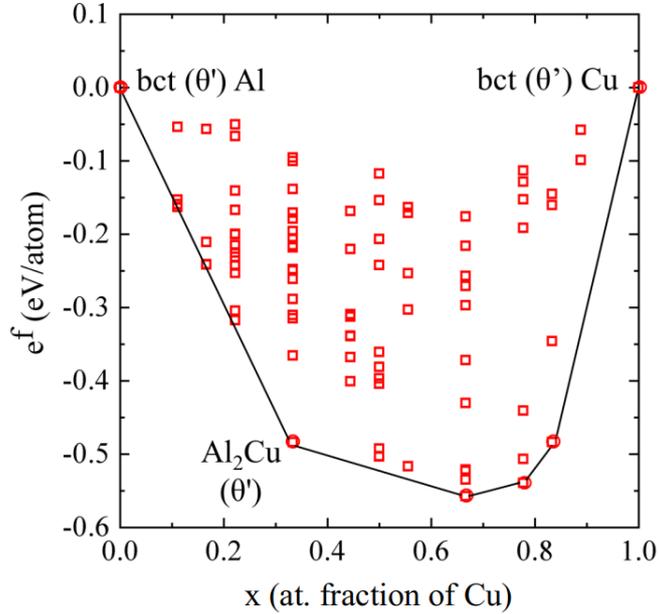

**Fig. 8** Formation energies of symmetrically distinct configurations in the bct (θ') Al-Cu system calculated by DFT with respect to the bct (θ') Al and Cu phases. The ground state phases in the convex hull are marked with a red circle.

The Gibbs free energies calculated from the CE of α-Al and θ' (using the formation energies of fcc Al and Cu as reference values) are shown in Fig. 9 at different temperatures (200K, 400K, 600K and 800K). The values of $G^{\theta'}$ at $x$=0.33 do not change with temperature, which means the θ' is a stochiometric line-compound. Thus, the vibrational entropy contribution of θ' was included, and the Gibbs free energies of α-Al and θ' including both configurational and vibrational entropic contributions are plotted in Fig. 10a at different temperatures. The phase boundaries can be determined as a function of temperature from the common tangent of the Gibbs free energies of α-Al and θ'', which is shown as a dashed black line in Fig. 10a for T = 800K as an example. The content of Cu in α-Al is determined by mapping the intersection of the Gibbs free energy of α-Al and the common tangent line onto the composition axis, while $x_{\theta'}$ = 0.33. It should be noted that the common tangent does not change even if you add the vibrational entropic contribution to the whole $G^{\theta'}$ curve.

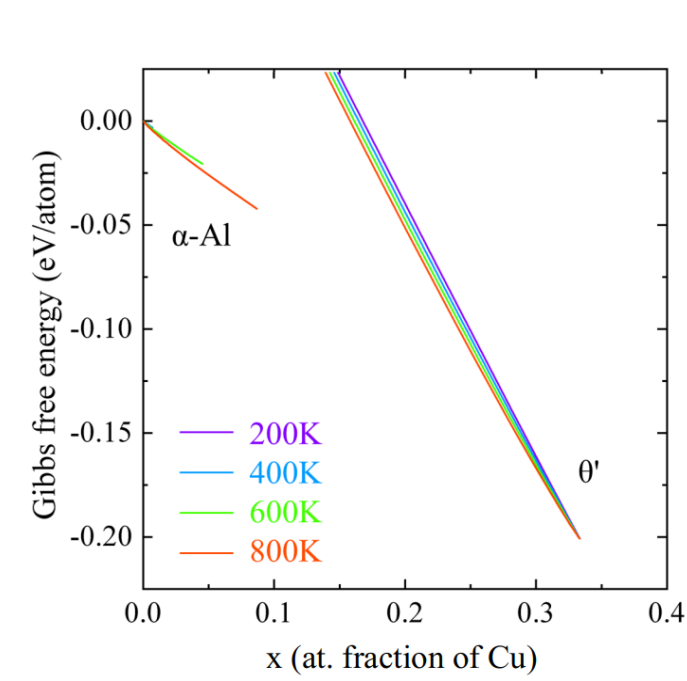

**Fig. 9** Calculated Gibbs free energies of α-Al and θ' as a function of composition *x* at different temperatures.

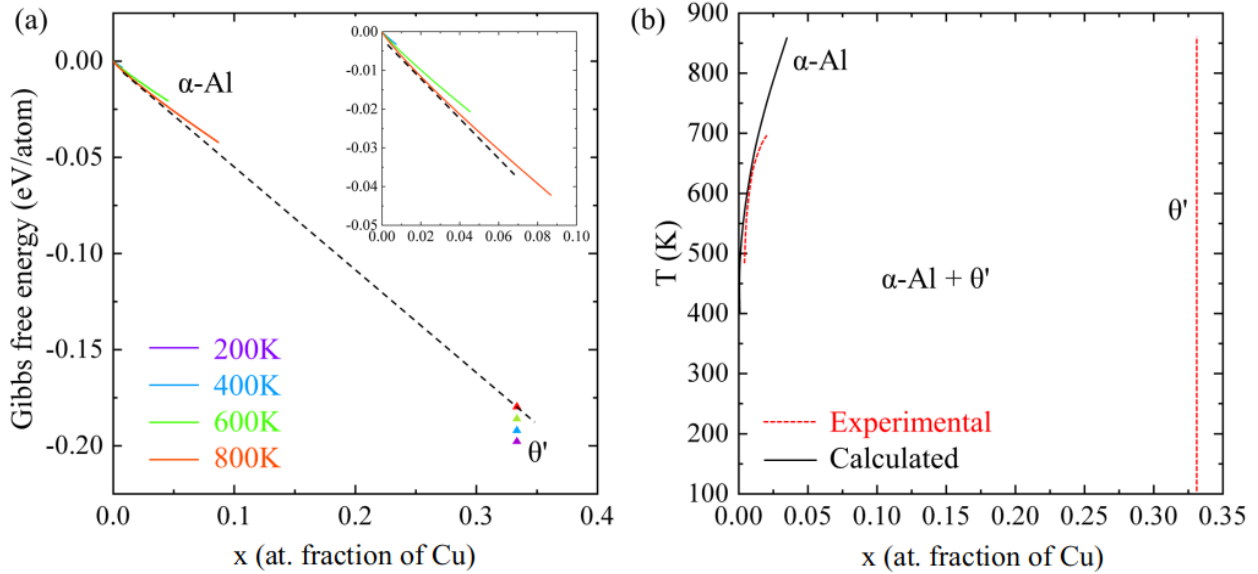

**Fig. 10** (a) Gibbs free energies of α-Al and θ' including both configurational and vibrational entropic contributions. The inset shows the detail of $G^\alpha$ at different temperatures. The differences in $G^{\theta'}$ with temperature are due to the contribution of vibrational entropy from [19]. The dashed black line shows the common tangent between $G^\alpha$ and $G^{\theta'}$ at 800K. (b) Phase boundaries between α-Al and θ'. The experimental curve corresponds to an average of different experimental results in [5].

This procedure can be repeated at different temperatures and the corresponding phase transition boundary between α-Al and θ' is shown in Fig. 10b, together with an average of the experimental data in the literature. The θ' solvus curve cannot be obtained by the hardness reversion method because it is very close to the equilibrium curve [5] and it was estimated by different authors from thermal effects during heating [41] or cooling [42] or resistivity measurements [43]. Nevertheless, these studies were not always accompanied by transmission electron microscopy observations, leading to uncertainties about the observed transitions. The data plotted in Fig. 10b correspond to the average solvus curve in [5] after a critical analysis of the experimental data. Again, they are in good agreement with the predictions obtained by CE, which are able to extend the phase transition range to a wider range of Cu composition. Obviously, both experimental and CE results indicate that θ' is a line compound at $x_{θ'} = 0.33$.

### 4.5 Phase boundary between α-Al and θ

The primitive cell of bct θ was used as the motif structure to get the thermodynamic properties of this phase. 93 symmetrically distinct configurations were generated with randomly arranged Al and Cu atoms on the lattice sites up to 12 atoms per unit cell. 7 of them over-relaxed to very different types of lattice and were not included to build up the CE. The formation energies of the 86 configurations used to create the CE are shown in Fig. 11. Clusters of atoms with maximum atomic spacing of 7 Å for pairs, 5 Å for triplets, and 3 Å for quadruplets were considered to determine the ECI coefficients of the CE for the bct (θ) Al-Cu system. A total of 90 clusters were included. The final optimized ECI coefficients set consists of a point cluster interaction, 8 pair cluster interactions and 10 triplet cluster interactions (see in Table. S3 in Supplemental Material). The corresponding cross-validation coefficient was 0.03 eV/atom.

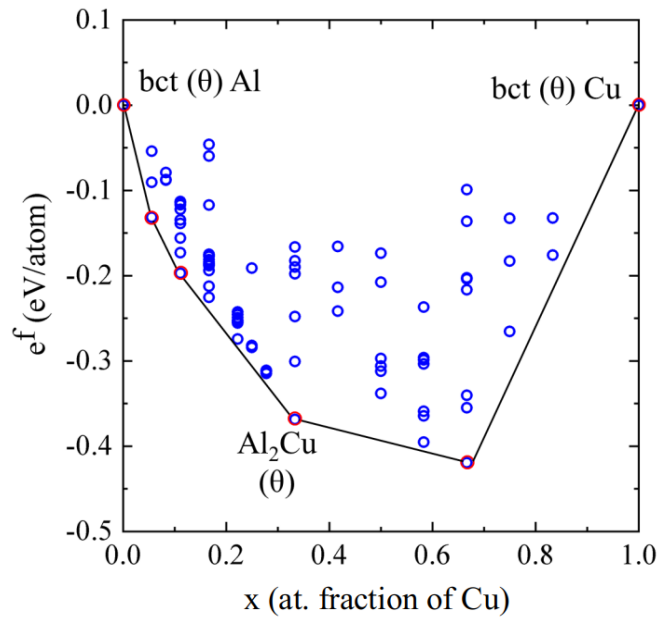

**Fig. 11** Formation energies of symmetrically distinct configurations in the bct (θ) Al-Cu system calculated by DFT with respect to the bct (θ) Al and Cu phases. The ground state phases in the convex hull are marked with a red circle.

The $G^\theta$ was obtained as a function of composition at different temperatures following the same procedure indicated in Section 4.4 where the reference values of $G^\theta$ were changed from those corresponding to the bct structures of Al and Cu to the reference values of fcc Al and Cu by adding the energy difference $\Delta E^f$ for each composition.

The Gibbs free energies of α-Al and θ are plotted in Fig. 12a as a function of temperature and the phase boundaries between α-Al and θ were obtained from the common tangent between $G^\alpha$ and $G^\theta$. It is plotted in Fig. 12b together with experimental data of the phase transition boundary [5]. The left boundary corresponding to the solubility limit of Cu in Al has been extensively characterized experimentally[1] and the CE results are in excellent agreement with the experimental data up to 821K, the eutectic temperature in the Al-rich region of the Al-Cu system. On the contrary, the experimental data on the phase boundary between α-Al and θ close to x = 0.33 reported in [5] are based in a few data points reported in an investigation carried out more than one century ago [44]. The CE simulations show significant differences and point out the experimental difficulties associated with the experimental determination of phase boundaries.

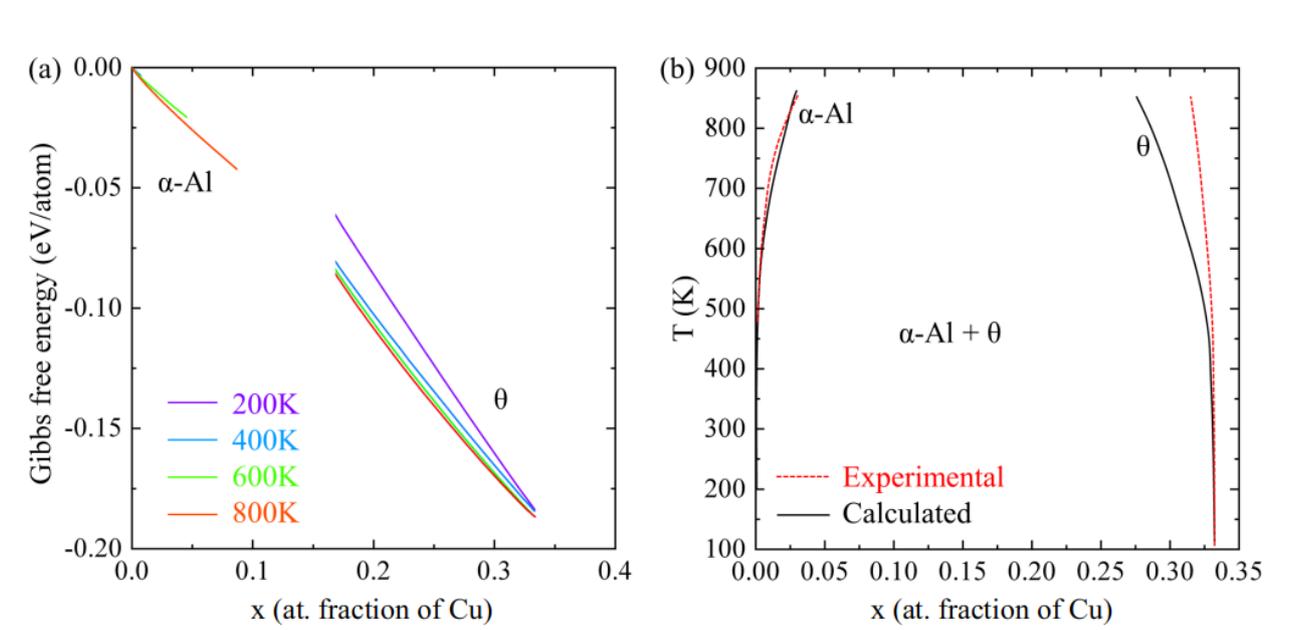

**Fig. 12** (a) Gibbs free energies of α-Al and θ as a function of composition for different temperatures. (b) Phase boundaries between α-Al and θ. The experimental curves correspond to an average of different experimental results in [5].

## 5. Discussion

### 5.1 GP zones and θ'' precipitates

The formation energies of all configurations in the three lattice structures obtained by DFT are plotted in Fig. 13. All of the them are referred to the formation energy of fcc Al and Cu. Up to $x=0.25$, the configurations of the fcc Al-Cu system that lie close to the segment connecting α-Al and θ" always present the lowest energies. Thus, the configurations lying near the line segment connecting α-Al and θ" are metastable phases that may be present at low temperatures and will disappear gradually with increasing temperature, the solution rate being controlled by the kinetics of the diffusion of Cu atoms in the fcc Al matrix.

---
[1] Data from 12 different experimental investigations are reported by Murray [5] and they practically superposed from 600K up to the eutectic temperature.

The structures of these configurations in Fig. 5 indicate that Cu atoms tend to be arranged on the {001} planes of the α-Al matrix because this structure has the lowest energy. This process will continue aggregating more Cu atoms, leading to the formation of a monolayer of Cu(001) embedded in the Al(001) layers, the so-called GP zones [45-46]. So, the GP zones are essentially periodic structures composed of Cu(001) monolayers embedded between Al(001) layers with different spacing between them. It is very likely that more configurations with Cu(001) monolayers embedded in Al(001) layers will appear on the line segment if more configurations are generated, and they would also be GP zones. Therefore, GP zones are not a well-established phase with a given composition but a range of phases formed by Cu(001) monolayers with different spacing between them that appear during the precipitation of θ".

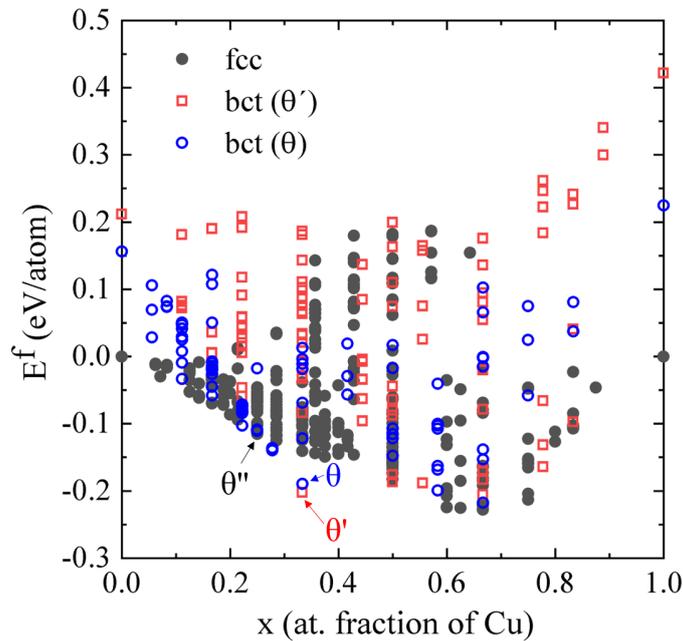

**Fig. 13** Formation energies obtained by DFT of all configurations in the three lattice structures: fcc, bct (θ') and bct (θ). The formation energies are referred to the fcc Al and Cu phases.

According to the structures shown in Fig. 5, the thickness of these GP zones is equal to one monolayer of Cu atoms, in agreement with experimental observations [7,31]. Growth of GP zones parallel to the disc depends on the diffusion of Cu atoms and also by the increase in interfacial energy as well as in strain energy as a result of the lattice mismatch between Cu(001) monolayers and the surrounding α-Al matrix. Note, however, that the calculated phase diagram is intended at the thermodynamics limit and, therefore, it does not consider the strain energy that might appear in multiphase regions.

The Gibbs free energies of α-Al and θ" are plotted in Fig. 14 as a function of temperature. The common tangent of Gibbs free energies of α-Al and θ" below 300K is very close to that of the GP zones with different arrangements indicated above. Nevertheless, common tangent between $G^{\alpha-Al}$ and $G^{\theta''}$ at temperatures higher than 300 K is always below that of the different GP zones with α-Al and the fcc lattice formed by three Al(001) layers sandwiched between two Cu(001) monolayers becomes the stable phase (Fig. 1b). Because the precipitation of θ" also depends on the diffusion of Cu atoms, it could be argued that the GP zones will be favorable sites for the nucleation of θ" because they already contain the basic blocks of the θ" precipitates, namely layers of Cu atoms on the {100} planes.

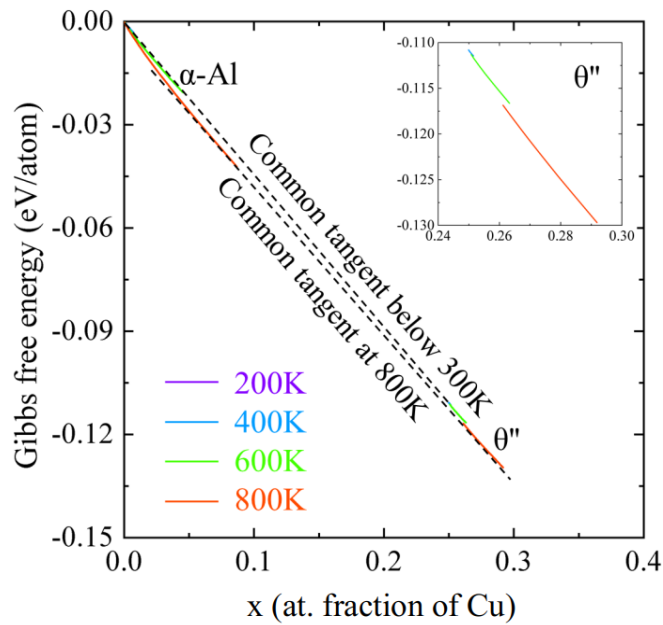

**Fig. 14** Gibbs free energies of α-Al and θ" at different temperatures. The inset shows the detail of $G^{\theta''}$. The dashed black lines show the common tangents between $G^{\alpha}$ and $G^{\theta''}$ at 800K and 300K.

It should be also noted that the Gibbs free energy curves of the $G^{\theta''}$ precipitates can be used to carry out parameter-free predictions of the growth of θ" precipitates at different temperatures through mesoscale phase field simulations [19]. In this analysis, the contributions from interface energy and elastic strain energy associated with the elastic mismatch between α-Al and θ" can also be accounted for from first principles simulations of the lattice parameters and elastic constants of both phases together with the interface energy of the coherent interfaces between both phases. The predictions of these phase field simulations were in good agreement with the equilibrium size and aspect ratio of the θ" precipitates after aging at 453K in an Al-1.7 at.% Cu alloy [19].

**5.2 θ' and θ precipitates**

According to the formation energies of different configurations in the fcc Al-Cu system (Fig. 3), the slopes connecting α-Al and θ" as well as θ" and $Al_2Cu_3$ are very close. Thus, the chemical potentials that stabilize α-Al and θ" as well as θ" and $Al_2Cu_3$ are very close. In fact, the chemical potentials that stabilize α-Al and θ" are gradually approaching those that stabilize θ" and $Al_2Cu_3$ as the temperature increases (Fig. 14). Nevertheless, neither $Al_2Cu_3$ nor AlCu are stable phases at low temperatures, because the bct θ' phase has lower formation energy (Fig. 13) and is on the convex hull. θ' is known to precipitate from the supersaturated solid solution after the θ" phase during aging at approximately < 473K and tends to nucleate along dislocations and grain boundaries [7, 31-32]. The θ' precipitates have a plate shape with $\{001\}_\alpha$ habit planes. The broad faces of the plates are nearly fully coherent with the α-Al matrix while the edges of the plates are semi-coherent. Precipitate growth has also been analyzed using the mesoscale phase field method and the habit plane and shape of these precipitates, as well as the trend to nucleate in dislocations comes about as a result of the interplay between the interface energy and the transformation strain associated to the nucleation of the precipitate [7, 18].

The Gibbs free energies of α-Al, θ' and θ are plotted in Fig. 15 as a function of temperature. While $G^\theta$ decreases with temperature, $G^{\theta'}$ increases with temperature due to the vibrational entropic contribution. As a result, θ' is the stable phase below 550K and it is replaced by θ above 550K, in agreement with previous analyses [19-20]. θ' and θ are made up of alternating Cu and Al monolayers and their stoichiometry is the same. Therefore, from the energy viewpoint, it seems feasible that θ' is transformed into θ at high temperature due to the entropic contribution although the transformation mechanisms are still not known. Once this transformation has taken place, θ may remain as a metastable phase if the alloy is rapidly quenched to low temperatures.

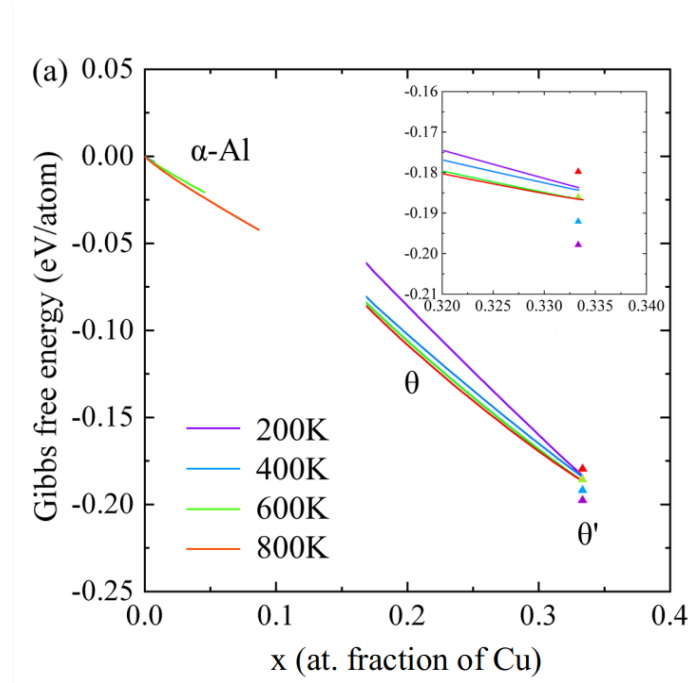

**Fig. 15** Gibbs free energies of α-Al, θ' and θ at different temperatures. The inset shows the detail of shows the detail of $G^{\theta'}$ and $G^{\theta}$ around x=0.33.

## 6. Conclusions

The thermodynamic properties of α-Al and other phases (GP zones, θ'', θ' and θ) in the Al-rich part of the Al-Cu system have been obtained by means of the cluster expansion formalism in combination with statistical mechanisms. In particular, the ground state phases in the convex hull were obtained by DFT calculations of phases with fcc, bct (θ') and bct (θ) lattice structures. Then, the grand potential and the Gibbs free energies of the ground state phases were determined by means of low thermal expansion and metropolis Monte Carlo simulations using the CASM code. Finally, the Al-rich part of the Al-Cu phase-diagram was built taking into account vibrational entropic contribution of the θ', as those of the other phases were negligible. The simulation predictions of the phase boundaries between α-Al and either θ'', θ' or θ phases as a function of temperature were in good agreement with the experimental data in the literature and expanded the experimental phase boundaries to a wider temperature range. They also showed that θ' is the stable phase below 550K but it is replaced by θ above this temperature because of the vibrational entropic contribution to the Gibbs energy of θ'. Finally, the DFT calculations showed the presence of a number of metastable configurations that may coexist with α-Al and θ'' at low temperatures. They are Guinier-Preston-zone type periodic structures composed of Cu(001) monolayers embedded between Al(001) layers with different spacing between them. The Gibbs free energy curves of the different phases obtained from cluster expansion can be used as input - in combination with interface and elastic mismatch energies - of precipitate nucleation and mesoscale phase field models of precipitate growth to make self-consistent predictions of precipitation in Al-Cu alloys.

**Acknowledgements**

This investigation was supported by the European Research Council (ERC) under the European Union's Horizon 2020 research and innovation programme (Advanced Grant VIRMETAL, grant agreement No. 669141). SL acknowledges the support from the Talent Attraction program of the Comunidad de Madrid (2018-T2/IND-10785). Computer resources and technical assistance provided by the Centro de Supercomputación y Visualización de Madrid (CeSViMa) are gratefully acknowledged. Finally, use of the computational resources of the Center for Nanoscale Materials, an Office of Science user facility, supported by the U.S. Department of Energy, Office of Science, Office of Basic Energy Sciences, under Contract No. DE-AC02-06CH11357, is also gratefully acknowledged. The support and advice of Dr. B. Puchala and Dr. I. Papadimitriou during the course of this work was much appreciated.

**References**

[1] T. M. Pollock, A. Van der Ven, The evolving landscape for alloy design. MRS Bulletin 44 (2019) 238-246.
[2] I. J. Beyerlein, S. Xu, J. LLorca, J. A. El-Awady, J. R. Mianroodi, B. Svendsen. Alloy design for mechanical properties: Conquering the length scales. MRS Bulletin 44 (2019) 257-265.
[3] H. L. Lukas, S. G. Fries, B. Sundman, Computational Thermodynamics, the Calphad Method. Cambridge University Press (2007).
[4] N. Ponweiser, C.L. Lengauer, K.W. Richter, Re-investigation of phase equilibria in the system Al-Cu and structural analysis of the high-temperature phase $\eta_1$-Al$_{1-\delta}$Cu, Intermetallics 19 (2011) 1737-1746.
[5] J. L. Murray, The aluminium-copper system. Int. Met. Rev. 30 (1995) 211-233.
[6] O. Zobac, A. Kroupa, A. Zemanova, K.W. Ritcher, Experimental description of the Al-Cu binary phase diagram, Metall. Mater. Trans. A 50 (2019) 3805-3815.
[7] J.F. Nie, "Physical metallurgy of light alloys," in Physical Metallurgy, chapter 20, pp. 2009–2156 Elsevier (2014).
[8] I. J. Polmear, Light Alloys, 2nd edition, Edward Arnold, London, 1989.
[9] J. M. Sanchez, Cluster expansions and the configurational energy of alloys, Phys. Rev. B 48 (1993) 14013.
[10] A. van de Valle, G. Ceder, Automating first-principles phase diagram calculations, J. Phase Equilib. 23 (2002) 348-359.
[11] A. Van der Ven, J.C. Thomas, B. Puchala, A. R. Natarajan, First-principles statistical mechanics of multicomponent crystals, Ann. Rev. Mater. Res. 48 (2018) 10.1–10.29.
[12] J. Teeriniemi, J. Huisman, P. Taskinen, K. Laasonen, First-principles modelling of solid Ni-Rh alloys, J. Alloy. Comp. 652 (2015) 371-378.
[13] J. Teeriniemi, M. Melander, S. Lipasti, R. Hatz, K. Laasonen, Fe-Ni nanoparticles: A multiscale first-principles study to predict geometry, structure, and catalytic activity, J. Phys. Chem. C 121 (2017) 1667-1674.
[14] J. Teeriniemi, P. Taskinen, K. Laasonen, First-principles investigation of the Cu-Ni, Cu-Pd, and Ni-Pd binary alloy systems, Intermetallics 57 (2015) 41-50.
[15] M. Asta, V. Ozolins, C. Woodward, A first-principles approach to modeling alloy phase equilibria, JOM 53 (2001) 16-19.
[16] A.R. Natarajan, E.L.S. Solomon, B. Puchala, A. Van der Ven, On the early stages of precipitation in dilute Mg-Nd alloys, Acta. Mater. 108 (2016) 367-379.
[17] N.A. Zarkevich, D.D. Johnson, Predicted hcp Ag-Al metastable phase diagram, equilibrium ground states, and precipitate structure. Phys. Rev. B 67 (2003) 064104.
[18] H. Liu, B. Bellón, J. LLorca, Multiscale modelling of the morphology and spatial distribution of θ' precipitates in Al-Cu alloys, Acta. Mater. 132 (2017) 611-626.
[19] H. Liu, I. Papadimitriou, F. X. Lin, J. LLorca, Precipitation during high temperature aging of Al-Cu alloys:

A multiscale analysis based on first principles calculations, Acta. Mater. 167 (2019) 121-135.

[20] C. Ravi, C. Wolverton, V. Ozolins, Predicting metastable phase boundaries in Al-Cu alloys from first-principles calculations of free energies: The role of atomic vibrations, Europhys. Lett. 73 (2006) 719-725.

[21] C. Wolverton, V. Ozolins, Entropically favored ordering: The metallurgy of $Al_2Cu$ revisited, Phys. Rev. Lett. 86 2001 5518

[22] A. van de Valle, M. Asta, Self-driven lattice-model Monte Carlo simulations alloy thermodynamic properties and phase diagrams, Modelling Simul. Mater. Sci. Eng. 10 (2002) 521–538.

[23] N. Metropolis, A.W. Rosenbluth, M.N. Rosenbluth, A.H. Teller, Equation of state calculations by fast computing machines, J. Chem. Phys. 21 (1953) 1087-1092.

[24] B. Puchala, A. Van der Ven, Thermodynamics of the Zr-O system from first-principles calculations, Phys. Rev. B 88 (2013) 094108.

[25] CASM, v0.2.1 (2017). Available from https://github.com/prisms-center/CASMcode. doi: 10.5281/zenodo.546148

[26] P Giannozzi, Quantum Espresso: a modular and open-source software project for quantum simulations of materials, J. Phys. Conden. Mater. 21 (2009) 395502.

[27] J. P. Perdew, Generalized gradient approximation made simple, Phys. Rev. Lett. 77 (1996) 3865.

[28] S. Baroni, P. Giannozzi, E. Isaev, Density-functional perturbation theory for quasi-harmonic calculations, Rev. Mineral. and Geochem. 71 (2010) 39-57.

[29] B. Montanari, N.M. Harrison, Lattice dynamics of $TiO_2$ rutile: influence of gradient corrections in density functional calculations, Chem. Phys. Lett. 364 (2002) 528-534.

[30] A. F. Kohan, P. D. Tepesch, G. Ceder, C. Wolverton, Computation of alloy phase diagrams at low temperatures, Comp. Mat. Sci. 9 (1998) 389-396.

[31] A. Rodríguez-Veiga, B. Bellón, I. Papadimitriou, G. Esteban-Manzanares, I. Sabirov, J. LLorca. J. Alloys Compounds, 757 (2018) 504-519.

[32] V. Vaithyanathan, C. Wolverton, L.Q. Chen, Multiscale modeling of θ' precipitation in Al-Cu binary alloys, Acta. Mater. 52 (2004) 2973-2987.

[33] E.V. Shalaev, N.I. Medvedev, First-principles study of the effect of iron on the crystal structure, stability and chemical bonding in the β-based AlCu ordered η2-phase and the pretransition state of a solid solution, Philos. Mag. 92 (2012) 1649-1662.

[34] T. Duguet, F. Senocq, L. Aloui, F. Haidara, D. Samelor, D. Mangelinck, C. Vahlas, Monitoring microstructure and phase transitions in thin films by high-temperature resistivity measurements, Surf. Interface. Anal. 44 (2012) 1162.

[35] M. Brokate, J. Sprekels, Hysteresis and Phase Transitions, Springer, 1996.

[36] R. H. Beton, E. C. Rollason, Hardness reversion. of dilute aluminium copper and aluminium. copper magnesium alloys, J. Inst. Met. 86 (1957-58) 77-85.

[37] K. G. Satyanarayana, K. Jayapalan, T. R. Ananthraman, A study of metastable equilibrium in Al-Cu alloys, Current Sci. 42 (1973) 6-9.

[38] V. Ozolins, M. Asta, Large vibrational effects upon calculated phase boundaries in Al-Sc, Phys. Rev. Lett. 86 (2001) 448-451.

[39] A. van de Walle, Multicomponent multisublattice alloys, nonconfigurational entropy and other additions to the Alloy Theoretic Automated Toolkit, CALPHAD 33 (2009) 266-278.

[40] K. Wang, D. Cheng, C.L. Fu, B.C. Zhou, First-principles investigation of the phase stability and early stages of precipitation in Mg-Sn alloys, Phys. Rev. Mater. 4 (2020) 013606.

[41] G. Borelius, J. Anderson, K. Guillberg, Ing. Vetenskaps Akad. Handl. 169 (1943) 5-27.

[42] H. Hori, K.Hirano, J. Jpn. Inst. Met. 37 (1973) 142-148.

[43] K. Matsuyama, Ternary diagram of the Al-Cu-Si system, Kinzoku-no Kenkyu, 11 (1934) 461–90.

[44] A. G. C. Gwyer, Über die legienrungen des aluminuims mit kupfer, eisen, nickel, kobalt, biei und cadmium. Z. Anorg. Chem. 57 (1908) 113-126.

[45] O.I. Gorbatov, Yu. N. Gornostyrev, P. A.Korzhavyi, Many-body mechanism of Guinier-Preston zones stabilization in Al–Cu alloys, Scripta Mater. 138 (2017) 130-133.

[46] H. Miyoshi, H. Kimizuka, A. Ishii, S. Ogata, Temperature-dependent nucleation kinetics of Guinier-Preston zones in Al-Cu alloys: an atomistic kinetic Monte Carlo and classical nucleation theory approach. Acta Mater. 179 (2019) 262-272.